\documentstyle[floats,aps,epsfig]{revtex}
\begin{document}
\def \beq{\begin{equation}}
\def \eeq{\end{equation}}
\def \beqarr{\begin{eqnarray}}
\def \eeqarr{\end{eqnarray}}
\def \be{\begin{equation}}
\def \ee{\end{equation}}
\def \bea{\begin{eqnarray}}
\def \eea{\end{eqnarray}}
\def \ta{{\tilde\alpha}}
\def \tg{{\tilde g}}     
\twocolumn[\hsize\textwidth\columnwidth\hsize\csname @twocolumnfalse\endcsname
\title{Critical Resistivity along the Quantum Hall Liquid--Insulator Transition
Line}
\author{Efrat Shimshoni}
\address{Department of Mathematics-Physics, Oranim--Haifa University,
Tivon 36006, Israel.}
\date{\today}
\maketitle
\begin{abstract}
The critical resistivity $\rho_{xx}^c$ measured along 
the quantum Hall liquid--insulator transition line indicates a pronounced 
peak for a critical filling factor $\nu_c\sim 1$. The origin of this behavior, 
which marks the crossover from the high to low magnetic field regime in the 
phase diagram, is explained in the framework of classical 
transport in a puddle network model. The proposed scenario is also 
consistent with the behavior of the critical Hall 
resistance along the transition line. In addition, a
functional form is suggested as a fit for isotherms $\rho_{xx}(\nu)$,
to be compared with experimental data in the moderately high field regime.  

\end{abstract}
\vskip2pc]
\narrowtext
The rich phase diagram of a disordered two--dimensional electron system in a 
perpendicular magnetic field consists of two prominent types of phases, 
distinguishable by their transport properties at low temperatures ($T$). These
include the quantum Hall (QH) liquid states, characterized by a quantized Hall
resistivity $\rho_{xy}$ accompanied by $\rho_{xx}\rightarrow 0$, 
and insulating states
in which  $\rho_{xx}\rightarrow \infty$. The transitions between adjacent
phases have been extensively studied \cite{phas-tran,dual,reslaw,qhi,hilke1}.
More recently, there has been much progress in the experimental effort 
to map the phase diagram in the integer QH regime \cite{hanein,hilke2}, and 
thus test the theoretically predicted structure \cite{thpd}. In particular,
a transition line separating the conducting phases from the insulator was 
identified in the $n$--$B$ plane (where $n$, the carrier density and $B$,
the magnetic field, are the two independent control parameters). Along 
this line, in the moderately low $B$ regime, direct transitions are observed
from $N>1$ QH states ($N$ integer) to the insulator \cite{hilke2}. This
behavior, and the overall structure of the diagram, are consistent with
a numerical result based on the tight binding model for non--interacting 
electrons \cite{SW}. 

Comparison of the transport data along the conductor--insulator transition 
line ($n_c(B_c)$) in Refs. \cite{hanein} and \cite{hilke2} 
indicates significant differences, particularly at low $B$ (where
the GaAs hole system studied in Ref. \cite{hanein} exhibits a $B=0$ 
metal--insulator transition at a finite $n_c$). These can be attributed to
sample dependent properties. However, the critical resistivity $\rho_{xx}^c$
measured along this transition line exhibits a striking common feature 
(observed in other samples as well \cite{Hilke}): it has a pronounced
peak at $\nu_c\sim 1$. Here $\nu_c$ is the critical value of the Landau level 
(LL) filling factor ($\nu\equiv n\phi_0/B$, with $\phi_0$ the flux quantum), 
so that $\nu_c\sim 1$ marks the crossover from high to low $B$, where in the 
latter, more than one LL is significantly occupied. This crossover is also
signified by the critical Hall resistivity $\rho_{xy}^c$ \cite{hilke2}:
for $\nu_c\leq 1$ it is quantized at $\rho_{xy}^c=h/e^2$, 
while above $\nu_c\sim 1$ it exhibits a classical Hall
effect $\rho_{xy}^c\approx h/\nu_ce^2$. 
$\rho_{xx}^c$ saturates to the universal value $\sim h/e^2$ in the high 
$B$ limit where $\nu_c\rightarrow 1/2$. However, the peak value of $\rho_{xx}^c$ is
significantly elevated from $h/e^2$ and sample dependent ($\sim 4h/e^2$ in
Ref. \cite{hanein}, $\sim 2h/e^2$ in Ref. \cite{hilke2}).

Since this puzzling behavior of $\rho_{xx}^c$ appears to be generic,
it calls for a simple explanation of its physical origin. In this paper I show 
that the observed dependence of $\rho_{xx}$ on $\nu$, and in particular the
peak of $\rho_{xx}^c$ at $\nu_c\sim 1$, follow from a phenomenological model 
for the transport, formerly introduced to interpret data in the high $B$ 
limit \cite{SA,SAK}. The essential assumption underlying this model is that
near the transition from a QH liquid to the insulator, the transport is 
dominated by narrow junctions connecting QH puddles; the puddles are defined 
by regions encircled by current carrying channels (edge states), and their 
typical size is assumed to be larger than a characteristic dephasing length.
As a result, the longitudinal resistance of the sample is that of a 
{\it classical} random resistor network, 
and the Hall resistance is quantized provided all
the QH puddles are at the same filling factor. This scenario successfully 
explains the experimentally observed quantized Hall insulator phenomenon 
\cite{qhi}, as well as the $\nu$--dependence of $\rho_{xx}$ \cite{reslaw}. More
recently, a similar model for the transport was shown to be consistent 
with the transport properties at low $B$, in samples indicating a $B=0$
metal--insulator transition \cite{Meir}. 

It is useful to start by showing that in the high $B$ limit, this model implies
$\rho_{xx}^c\approx h/e^2$, regardless of the details of the disorder 
potential. In this limit only the lowest LL is occupied, 
hence all the QH puddles contain a single conducting channel. The longitudinal
resistance of the junction $j$ between adjacent puddles in the network is 
therefore given by the Landauer--B{\"u}ttiker formula \cite{LB} 
\beq
R_{xx}^j={h\over e^2}{{\cal R}_j\over {\cal T}_j}\; ,
\label{rland}
\eeq
where ${\cal T}_j=1-{\cal R}_j$ is the transmission probability across the
junction; the Hall resistance $R_{xy}^j=h/e^2$ for {\it any} $j$. As proven
in Ref. \cite{SA}, the global resistivity tensor of the sample conveniently
separates into Hall and longitudinal components: $\rho_{xy}=h/e^2$, and
$\rho_{xx}$ is set by the random network of the resistors $\{R_{xx}^j\}$. 
Namely, $\rho_{xx}\approx R_{xx}^p$, $p$ being the resistor at the percolation 
threshold of the distribution \cite{AHL}. The dependence of $\rho_{xx}$ on 
$T$ and $\nu$ is therefore determined by ${\cal T}_p(T,\nu)$.
For sufficiently high $T$ such that quantum tunneling is neglected 
\cite{Meir},
\beq
{\cal T}_p(T,\nu)=f(\epsilon_p-\epsilon_F(\nu))\; ,\quad 
f(\epsilon)\equiv {1\over 1+e^{\epsilon/k_BT}}\; ;
\label{Tp}
\eeq   
here $\epsilon_F(\nu)$ is the Fermi level, and $\epsilon_p$ is the energy of 
the saddle point at which the junction $p$ is located. Employing Eq. 
(\ref{rland}), one obtains
\beq
\rho_{xx}\approx{h\over e^2}{1-{\cal T}_p\over {\cal T}_p}
={h\over e^2}\exp\left[{\epsilon_p-\epsilon_F(\nu)\over k_BT}\right]\; .
\label{hBres}
\eeq    
It follows immediately that the critical filling factor $\nu_c$, defined such
that $\rho_{xx}(\nu_c)$ is {\it independent of} $T$, obeys 
$\epsilon_F(\nu_c)=\epsilon_p$ so that the exponent in Eq. (\ref{hBres})
{\it vanishes}. Consequently,
\beq
\rho_{xx}^c=\rho_{xx}(\nu_c)\approx{h\over e^2}\; .
\eeq

Note that this argument can be extended to account for quantum tunneling across
the junctions. Eq. (\ref{Tp}) is then generalized to a Sommerfeld expansion 
of the form
\beq
{\cal T}_p(T,\nu)=f(\epsilon_p-\epsilon_F(\nu))+
\sum_{n=1}^\infty \alpha_n(T)f^{(2n)}(\epsilon_p-\epsilon_F(\nu))\; ,
\eeq
where $f^{(2n)}(x)$ is the $2n$-th derivative of the Fermi function. This yield
a $T$--independent expression for $R_{xx}^p$ provided all terms in the sum
vanish. Since $f^{(2n)}(x)=-f^{(2n)}(-x)$, this occurs for 
$\epsilon_F(\nu)=\epsilon_p$, and once again $\rho_{xx}^c\approx h/e^2$. It is
also important to notice that the above argument for the universality of
$\rho_{xx}^c$ is not violated in the case where $\nu_c$ is different from 
$1/2$ due to lack of particle--hole symmetry in the disorder potential. The 
only essential assumption is the neglect of mixing with higher LL.
An interesting example for a remarkable universality of $\rho_{xx}^c$ at 
arbitrary $\nu_c$ is provided by the transitions from a {\it fractional} QH 
liquid to the insulator \cite{dual}; this observation will be discussed 
towards the end of the paper.    

The above scenario changes once the magnetic field $B$ (at a given $n$)
is reduced to a point where states in the second LL are occupied. In terms
of the configuration of conducting channels in real space, this corresponds
to the formation of closed loops of a second channel at the Fermi level inside
the QH puddles. The overall filling factor of the system is given by
\beq
\nu=p[(1-p_{h})+\nu_hp_{h}]\; ,
\label{nup}
\eeq
where $p\equiv A_l/A$ is the total area fraction of the conducting liquid,
$p_{h}\equiv A_{h}/A_l$ is the relative area fraction occupies
by higher LL states and $\nu_h\geq 2$ their local filling factor.
For a smoothly varrying disorder potential, $p_{h}$
can be estimated noting that $p_{h}=r_2^2/r_1^2$, where $r_{1(2)}$ is
the average radius of the region surrounded by the first (second) LL
channel in a typical QH puddle, and
\beq
{1\over 2}Kr_1^2\approx{1\over 2}Kr_2^2+\Delta(B)\; ;
\label{r1r2}
\eeq
$\Delta(B)$ is either the cyclotron gap $\hbar\omega_c$ or the Zeeman
energy gap for spin--resolved Landau bands, and $K$ is the average curvature of
the potential local minima. The critical filling factor $\nu_c$ for a transition to the
insulator is then expressed as
\beq
\nu_c=p_c\left[1+{(\nu_h-1)r_2^2\over r_1^2}\right]\approx 
p_c\left[\nu_h- {B_c\over B_0}\right]\; ,
\label{nucB}
\eeq
where $p_c$ is the percolation threshold of the liquid, $B_c$ the critical
field, and $B_0$ is a sample dependent parameter: 
$B_0=m^\ast cKA_p/he(\nu_h-1)$, $B_0=KA_p/2\pi g\mu_B(\nu_h-1)$ for the 
spin--unresolved and spin--resolved cases, respectively ($m^\ast$ 
is the effective mass of the carriers, $A_p$ is
the typical puddle area at percolation). Note that Eq. (\ref{nucB}) is valid
for $0\leq r_2\leq r_1$, i.e. for $p_c\leq \nu_c\leq \nu_hp_c$; 
for a particle--hole symmetric potential $p_c=1/2$, hence the expression 
describes an interpolation between $\nu_c\sim 1/2$ (at high $B$) and 
$\nu_c\sim \nu_h/2$. Neglecting the occupation of LL with $N>2$, $\nu_h\sim 2$
and the upper limit of this regime is $\nu_c\sim 1$. The critical
density is then given by
\beq
n_c\approx {1\over \phi_0}\left[B_c-{B_c^2\over 2B_0}\right]\; ,
\label{ncvsBc}
\eeq
which is indeed a reasonable fit to the shape of the critical line $n_c(B_c)$
in Ref. \cite{hanein}, for $\nu_c<1$.

\begin{figure}[htb]
\vskip0.5in
\centerline{\epsfig{file=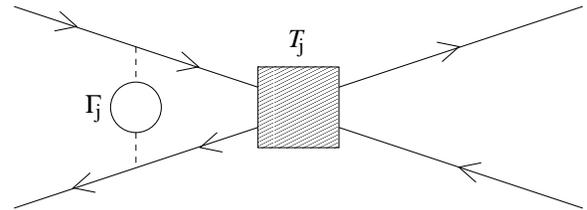,width=3in,angle=-0}}
\vskip0.5in
\caption[]
{
\label{fig:fig1}
The configuration of conducting channels near a typical junction connecting 
adjacent QH puddles.
}
\end{figure}

More interesting is the effect of occupied higher LL states on $\rho_{xx}$
as $\nu_c\rightarrow 1$. In the regime $\nu_c<1$, corresponding to
$p_h<1$, these states are localized, forming closed loops inside
the QH puddles. A typical junction in the network of conducting channels
is schematically depicted in Fig. 1. As $p_h\rightarrow 1$,
the localized high LL states are coupled to the primary conducting channels 
via tunneling through occasional weak links (dashed
lines in the figure), and thus play the role of resonant scatterers
which induce backscattering across the conducting regions. Denoting this
backscattering rate close to the $j$-th junction by $\Gamma_j$, and
neglecting quantum interference between the two reflection paths
\cite{fn1}, the effective transmission through the junction is reduced by a 
factor of $(1-\Gamma_j)$. The junction resistance becomes
\beq
R_{xx}^j={h\over e^2}{1-(1-\Gamma_j){\cal T}_j\over (1-\Gamma_j){\cal T}_j}\; .
\label{rlandmod}
\eeq
The entire distribution of resistors in the random network is thus shifted
up, and consequently the network resistance (determined as usual by the percolative
resistor $R_{xx}^p$) is  
\beq
\rho_{xx}\approx{h\over e^2}{1-(1-\Gamma_p)f(\epsilon_p-\epsilon_F(\nu))\over
(1-\Gamma_p)f(\epsilon_p-\epsilon_F(\nu))}\; .
\label{intBres}
\eeq
Here I assumed that the dependence of $\Gamma_p$ on $T$ is
negligible compared to the transmission across the junction, which
involves an energy barrier. Eq. (\ref{intBres}) clearly implies that
the $T$ dependence disappears at a single value of the filling factor
$\nu_c$ where $\epsilon_F(\nu_c)=\epsilon_p$, as in the high
$B$ limit. However, the value of the critical resistivity thus defined
is modified:
\beq
\rho_{xx}^c\approx{h\over e^2}{1+\Gamma_p\over 1-\Gamma_p}>{h\over e^2}\; .
\label{inthBrc}
\eeq
For $\nu_c\leq 1$, the area fraction $p_h$ increases, implying that
$\Gamma_p$ is obviously an increasing function of $\nu_c$ (the explicite
behavior of $\Gamma_p(\nu_c)$ depends on details of the disorder
potential). As a result, $\rho_{xx}^c$ {\it grows} with $\nu_c$ towards
the limit case $\nu_c\sim 1$.

Another characteristic of the $\nu_c<1$ (high $B$) regime is that in 
the entire range of parameters where the classical transport scenario is 
valid, the Hall resistance maintains the quantized value $\rho_{xy}=h/e^2$.
This follows from the fact that the Hall resistance of each junction as
depicted in Fig. 1 is determined by the chemical potential difference between
parallel right and left moving conducting channels, which is not affected by 
the modified transmission probability and yields $R_{xy}^j=h/e^2$ for any $j$.

The case where $\nu_c\sim 1$ marks a crossover to a low $B$ regime. 
From Eq. (\ref{nucB}) (assuming $p_c\approx 1/2$, $\nu_h\approx 2$), 
$\nu_c=1$ corresponds to $r_2\approx r_1$, which means that the average 
size of regions with a local filling fraction $\nu\geq 2$ coincides with the 
average size of the conducting 
regions \cite{fn2}. Beyond this threshold value of $\nu_c$, higher LL 
states are delocalized, adding conducting channels to the percolating 
(or nearly percolating) network. As a consequence, the resistance of a
typical junction $R_{xx}^j$ acquires a form similar to Eq. (\ref{rlandmod}),
reduced by an overall factor of $1/N$, where $N$ is the number of conducting
channels that are clamped together. 
Note that $\Gamma_j$ then describing scattering to partially 
occupied LL higher then $N$. The resulting critical value of the global 
longitudinal resistivity is then {\it reduced} compared to its value at  
$\nu_c\sim 1$, and is given by an expression of the form
\beq
\rho_{xx}^c\approx{h\over Ne^2}{1+\Gamma_p\over 1-\Gamma_p}\; .
\label{intlBrc}
\eeq
Here both $N$ and $\Gamma_p$ depend on $\nu_c$ in a complicated manner,
hence the actual behavior of $\rho_{xx}^c$ at $\nu_c\gg 1$ ($B\rightarrow 0)$ 
is expected to be strongly sample dependent. Indeed, in Ref. \cite{hilke2}
$\rho_{xx}^c\rightarrow 0$ in this limit, while in Ref. \cite{hanein} it 
saturates back to $\rho_{xx}^c\sim h/e^2$. However, in proximity to the 
threshold $\nu_c\sim 1$ the generic trend is a reduction of $\rho_{xx}^c$
when $\nu_c$ is increased beyond unity. It therefore follows that this
crossover between the high and low $B$ regimes is indicated by a peak in
the critical resistance.

The increasing number of conducting channels `clamped' together 
in the low $B$ regime also 
modifies the Hall resistance, since $\{R_{xy}^j\}$ are no longer identical.
As argued in Ref. \cite{SA}, the pure transverse component of the resistivity
measured across the sample can be expressed as
\beq
\rho_{xy}={h\over e^2}{\sum_i(I_i/N_i)\over \sum_iI_i}\; ,
\label{RxylB}
\eeq  
where $I_i$ is the local current through the $i$-th QH puddle between the
voltage probes, and $N_i$ its local filling factor. This yields a quantized 
value only provided all the $N_i$'s are identical. 
This is no longer the case for 
$\nu>1$: Eq. (\ref{RxylB}) generally describes a weighted average in which
the weight of terms with $N_i>1$ is monotonically increasing with $\nu$. As a
result $\rho_{xy}\sim h/\nu e^2$, as in the classical Hall regime. Along the
critical line, this implies $\rho_{xy}^c\sim h/\nu_c e^2$. The transition 
points in the vicinity of integer values $N>1$ of $\nu_c$ correspond to the 
cases where a majority of the QH puddles consists of a unique filling factor 
$N$. They therefore characterize direct transitions from $N>1$ QH phases to
the insulator.

The role of mixing with high LL does not only lead to violation of the
universality of $\rho_{xx}^c$: naturally, the behavior of $\rho_{xx}(\nu,T)$
across the transition is also modified. Indeed, in the high $B$ limit, 
isotherms $\rho_{xx}$ vs. $\nu$ can be fitted to a simple formula \cite{reslaw}
\beq
\rho_{xx}={h\over e^2}\exp\left[{-\Delta\nu\over \nu_0(T)}\right]\; ,
\label{rlaw}
\eeq 
where $\Delta\nu=\nu-\nu_c$ and $\nu_0(T)\approx\alpha T+\beta$. This formula 
exhibits a duality symmetry $\rho_{xx}(\Delta\nu)=\rho_{xx}^{-1}(-\Delta\nu)$,
which is eventually violated in the lower $B$ regime. The analysis presented 
in this paper implies a specific modification of the above formula, which
can be used as a test for the proposed transport scenario.
As shown in Ref. \cite{SAK}, the simple exponential dependence on 
$\Delta\nu$ in Eq. (\ref{rlaw}) can be derived from the energy dependent
expression Eq. (\ref{hBres}) (with an effective activation temperature) 
assuming a parabolic shape of the barriers, such that 
$(\epsilon_p-\epsilon_F(\nu))\propto(-\Delta\nu)$. Employing the same
approximation in the moderately high $B$ regime, where the resistivity is
given by Eq. (\ref{intBres}), and relating the parameter $\Gamma_p$ to 
$\rho_{xx}^c$ through Eq. (\ref{inthBrc}), the isotherms acquire the form 
\beq
\rho_{xx}(\nu)\approx {(\rho_{xx}^c-h/e^2)\over 2}+
{(\rho_{xx}^c+h/e^2)\over 2}\exp\left[{-\Delta\nu\over \nu_0(T)}\right]\; .
\label{intBrlaw}
\eeq
In Fig. 2, a few such isotherms are plotted for a case where $\rho_{xx}^c$ is 
considerably elevated from $h/e^2$. Note that for $\Delta\nu>0$, 
$\log\rho_{xx}$ vs. $\Delta\nu$ curve upward compared to the straight line 
expected from the symmetric formula Eq. (\ref{rlaw}). A preliminary comparison
with experimental data (e.g. Ref. \cite{hilke1}) confirms this behavior.

\begin{figure}[htb]
\vskip0.5in
\centerline{\epsfig{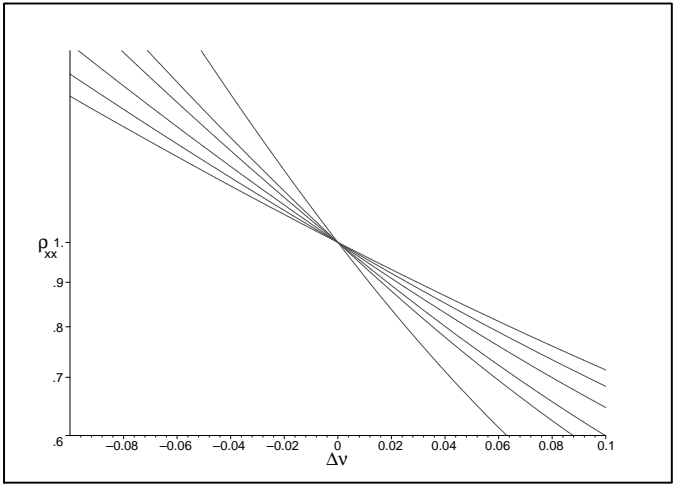}}
\vskip0.5in
\caption[]
{
\label{fig:fig2}
The longitudinal resistivity (normalized in units of $\rho_{xx}^c$) on a
log scale as a function of filling factor, computed from Eq. (\ref{intBrlaw})
for $\rho_{xx}^c=2.2h/e^2$, and $\nu_0=0.079,0.11,0.125,0.15,0.175,0.2$ 
($\nu_0=0.079$ is the steepest curve). 
}
\end{figure}

Finally, it is interesting to comment on the application of the transport
model described here for transitions from the fractional QH state $1/3$ to the
insulator. Experimentally \cite{dual}, the transport data exhibit the same
behavior as near the integer transitions: e.g., Eq. (\ref{rlaw}) is obeyed
replacing $\nu$ by an effective filling factor of composite Fermions. This
suggests that a similar model describes here the d.c. transport properties
of composite Fermions. In particular, high LL of Fermions correspond to 
the hirarchy fractions $2/5,3/7$ etc. In analogy with the integer case, 
occupation of localized high LL Fermion states would imply the formation of
incompressible regions in the disordered
sample at these filling fractions. However, in
contrast with the integer case, the energy gap supporting such regions is 
tiny. As a result, a mechanism for enhancing the resistance by backscattering
through localized high LL states is expected to be practically absent. This
possibly explains the observation that the duality symmetry in $\rho_{xx}$, as
well as the universality of $\rho_{xx}^c$ that comes with it, are better
manifested by data near the fractional transitions.    


I thank A. Auerbach, M. Hilke, A. Kamenev, D. Shahar, and S. Sondhi for useful 
conversations. This work was supported by grant no. 96--00294 from the 
United States--Israel Binational Science Foundation (BSF), Jerusalem, Israel.


\begin{references}
\bibitem{phas-tran}
For a review and extensive references, see A.~M.~M. Pruisken in 
{\it The Quantum Hall Effect,} Eds. R. E. Prange and S. M. Girvin
(Springer-Verlag, New York, 1986); S. Das Sarma in {\it Perspectives in the 
Quantum Hall Effect}, Eds. S. Das Sarma and A. Pinczuk (John Wiley and
Sons, 1997); S. L. Sondhi, S. M. Girvin, J. P. Carini and D. Shahar, 
Rev. Mod. Phys. {\bf 69}, 315 (1997).
\bibitem{dual}
D. Shahar, D. C. Tsui, M. Shayegan, R. N. Bhatt and J. E. Cunningham, Phys. 
Rev. Lett. {\bf 74}, 4511 (1995);
D. Shahar, D. C. Tsui, M. Shayegan, E. Shimshoni and S. L. Sondhi, 
Science {\bf 274}, 589 (1996).
\bibitem{reslaw}
D. Shahar, M. Hilke, C. C. Li, D. C. Tsui, S. L. Sondhi and M. Razeghi,
Solid State Comm. {\bf 107}, 19 (1998).
\bibitem{qhi}
D. Shahar, D. C. Tsui, M. Shayegan,
J. E. Cunningham, E. Shimshoni an d S. L. Sondhi, Solid State Comm. {\bf 102},
817 (1997); M. Hilke, D. Shahar, S. H. Song,  D. C. Tsui, Y. H. Xie and Don
Monroe, Nature  {\bf 395}, 673 (1998);
M. V. Yakunin, Yu. G. Arapov, O. A. Kuznetsov and V. N. Neverov, 
JETP Lett. {\bf 70}, 301 (1999).
\bibitem{hilke1}
M. Hilke, D. Shahar, S. H. Song,  D. C. Tsui, Y. H. Xie and Don Monroe, Phys. 
Rev. B {\bf 56}, R15525 (1997).
\bibitem{hanein}
Y. Hanein, D. Shahar, H. Shtrikman, J. Yoon, C. C. Li and D. C. Tsui,
Nature  {\bf 400}, 735 (1999).
\bibitem{hilke2}
M. Hilke, D. Shahar, S. H. Song,  D. C. Tsui and Y. H. Xie, 
to be published in Phys. Rev. B (preprint cond--mat/9906212).
\bibitem{thpd}
D. E. Khmel'nitzkii, JETP Lett. {\bf 38}, 552 (1983); 
S. Kivelson, D. H. Lee and S. C. Zhang, Phys. Rev. B {\bf 46}, 2223 (1992).
\bibitem{SW}
D. N. Sheng and Z. Y. Weng, cond--mat/9906261.
\bibitem{Hilke}
M. Hilke, private communication.
\bibitem{SA}
E. Shimshoni and A. Auerbach, Phys. Rev. B {\bf 55}, 9817 (1997).
\bibitem{SAK}
E. Shimshoni, A. Auerbach and A. Kapitulnik,  Phys. Rev. Lett. {\bf 80}, 
3352 (1998).
\bibitem{Meir}
Y. Meir, Phys. Rev. Lett. {\bf 83}, 3506 (1999); Y. Meir,
cond--mat/9912423.
\bibitem{LB}
M. B{\"u}ttiker, Phys. Rev. Lett. {\bf 57}, 1761 (1986); M. B{\"u}ttiker,
Phys. Rev. B {\bf 38}, 9375 (1988).
\bibitem{AHL}
V. Ambegaokar, B. I. Halperin and J. S. Langer, Phys. Rev. B {\bf 4}, 2612
(1971).
\bibitem{fn1}
The weak links to higher LL channels are located at 
arbirary distances from the nearest junctions in the network, which are
typically of the same order as the size of a puddle.
\bibitem{fn2}
The actual crossover value of $\nu_c$ is slightly higher 
than 1 for $\nu_h>2$; see, e.g., Ref. \cite{hilke2}.
\end{references}
\end{document}